\documentclass[aps,prl,twocolumn,superscriptaddress,longbibliographys]{revtex4-2}
\usepackage{graphicx}
\usepackage{bm}
\usepackage{natbib}
\usepackage{color}
\usepackage{epstopdf}
\usepackage{amsmath} 
\usepackage{url}
\usepackage{braket}
\definecolor{darkGreen}{RGB}{0,110,0}
\definecolor{darkBlue}{RGB}{0,0,130}
\usepackage[colorlinks=true, urlcolor=darkBlue,citecolor=darkGreen,linkcolor=darkBlue]{hyperref}
\begin{document}

\title{Local dynamics and detection of topology in spin-1 chains}

\author{Alfonso Maiellaro}
\affiliation {Universit\'e Paris-Saclay, CNRS, Laboratoire de Physique des Solides, 91405 Orsay, France}
\affiliation {INFN, Sezione di Napoli, Gruppo collegato di Salerno, Italy}
\author{Herv\'e Aubin}
\affiliation {Universit\'e Paris-Saclay, CNRS, Centre de Nanosciences et de Nanotechnologies, 91120, Palaiseau, France}
\author{Andrej Mesaros}
\email{andrej.mesaros@universite-paris-saclay.fr}
\author{Pascal Simon}
\email{pascal.simon@universite-paris-saclay.fr}
\affiliation {Universit\'e Paris-Saclay, CNRS, Laboratoire de Physique des Solides, 91405 Orsay, France}

\date{\today}

\begin{abstract}
Antiferromagnetic spin-1 chains host the celebrated symmetry protected topological Haldane phase, whose spin-$1/2$ edge states were evidenced in bulk by, e.g., Electron Spin Resonance (ESR). Recent success in assembling effective spin-1 antiferromagnetic chains from nanographene and porphyrin molecules opens the possibility of local, site-by-site, characterization. 
The nascent technique of combined ESR-STM is able to measure the spin dynamics with atomic real-space resolution, and could fully reveal and manipulate the spin-$1/2$ degree of freedom. 
In this work, we combine exact diagonalization and DMRG to investigate the local dynamic spin structure factor of the different phases of the bilinear-biquadratic Hamiltonian with single-ion anisotropy in presence of an external magnetic field. We find that the signature of the Haldane phase is a low-energy peak created by singlet-triplet transitions in the edge-state manifold. We predict that the signature peak is experimentally observable, although for chains of length above $N=30$ its energy should be first tuned by application of external magnetic field. We fully characterize the peak in real-space and energy, and further show its robustness to weak anisotropy and a relevant range of temperatures.
\end{abstract}

\maketitle

\paragraph*{Introduction.---} In accordance with Haldane's conjecture\cite{HalConj} which states that integer-spin antiferromagnetic chains have an energy gap, a non-degenerate ground state of a closed chain with a finite excitation gap was confirmed by the exact AKLT solution\cite{AKLT87,AKLT88} of a particular biquadratic antiferromagnetic spin-$1$ chain Hamiltonian. The "Haldane phase" of spin-$1$ chains\cite{Tasaki,RenardRev} adiabatically connected to the AKLT state, as any interaction driven insulator in one dimension, has hidden non-local order\cite{vanNijs,Kennedy1992,Montorsi}, and is today understood as a symmetry-protected topological phase in one dimension\cite{Wen2011,Pollmann2012}. The topology explains\cite{PollmannLesHouches} the early realization\cite{AKLT87,AKLT88} that an open chain in the thermodynamic limit hosts a projective, spin-$1/2$, degree of freedom on each edge\cite{Polizzi}, making the ground state four-fold (quasi)degenerate within the sectors of total spin $S=0$ and $S=1$. Experimentally, the presence of free spin-$1/2$ degrees of freedom was evidenced in quasi-1d antiferromagnetic bulk compounds using ESR and NMR (see Ref.\cite{RenardRev} for a review). More recently, exciting progress in direct access to the Haldane phase was made by engineering effective antiferromagnetic spin-$1$ chains from open-shell nanographene\cite{Mishra2021} and phtalocyanine\cite{Zhao2023}, where the expectation value of spin seems to indicate a spin-$1/2$ edge state. Theoretically, since the Haldane phase is realized due to interactions, the study of static properties such as magnetization of edge states\cite{RenardRev,Polizzi,Mishra2021}, single-particle excitations and static susceptibility\cite{Oitmaa09}, typically includes heavy numerics calculations. Understanding the spin dynamics through dynamic susceptibility is an even bigger challenge, recently addressed using translational symmetry of closed chains\cite{Fehske2018,Yao2021}, but remains largely unexplored for open chains.

In recent years a new experimental technique combining Scanning Tunneling Microscopy (STM) with Electron Spin Resonance (ESR)\cite{Yang2018,Willke2018,Zhang2022} revealed local spin dynamics with atomic precision. As the technique has progressed from isolated atoms\cite{Yang2018,Willke2018,Wang2023} to molecules\cite{Willke2021,Zhang2022,Zhang2023}, there is clear motivation to understand the local atom-by-atom spin dynamics of the edge states of a chain in a Haldane phase, which is known to constitute a resource for universal measurement-based quantum computation\cite{Wei2012}, and may provide longer spin coherence times due to topological protection. The existing regular STM measurements, in accord with numerical modeling\cite{Mishra2021}, find that starting from the edge of the chain the static magnetization decays exponentially, as expected universally due to the energy gap. Additionally, the magnetization oscillates, hence there are some non-universal features depending on the details of the Hamiltonian. Notably, with most chains being of lengths less than $N=16$, the observed magnetization decays too slowly for the bulk of the chain to be well seen\cite{Mishra2021,Zhao2023}.

In this Letter, we theoretically study the local spin dynamics in spin-1 chains, which may be accessed by ESR-STM, to reveal how 
an edge state behaves as an independent spin-$1/2$, and to better characterize the Haldane phase in shorter chains relevant for experiments.
%
%
More specifically, we focus on the dynamical spin structure factor (DSF) as the fundamental observable of spin dynamics. Using both exact diagonalization and DMRG we calculate the DSF atom-by-atom along a chain, for a large range of chain lengths $N=6\ldots 30$. We consider a family of spin-$1$ Hamiltonians that realize several phases, such as antiferromagnetic, ferromagnetic, and the Haldane phase. Our main finding is that DSF peaks locally at the edge of chain for a certain energy, but uniquely in the Haldane phase, thanks to transitions between the $S=0$ and $S=1$ components of the edge states. We also find that this behavior should be practically useful for characterization of short and medium chains, while for longer chains applying an external magnetic field at low temperatures simply increases the energy of the peak and makes it observable. We also find that the spatial profile of the DSF in the Haldane phase indicates the proximity of a transition to different trivial gapped states, making it plausible to extract the Hamiltonian parameters from simple observations of the local dynamics. Somewhat surprisingly, the signature peak in DSF is lost for the special case of AKLT state due to its finely tuned exact degeneracy of the edge states for any chain length, but the signal is strong for the Heisenberg antiferromagnet. In summary, we theoretically predict that local spin dynamics of experimentally available spin chains would powerfully characterize and distinguish the trivial and topological phases.

\begin{figure*}
	\includegraphics[width=\textwidth]{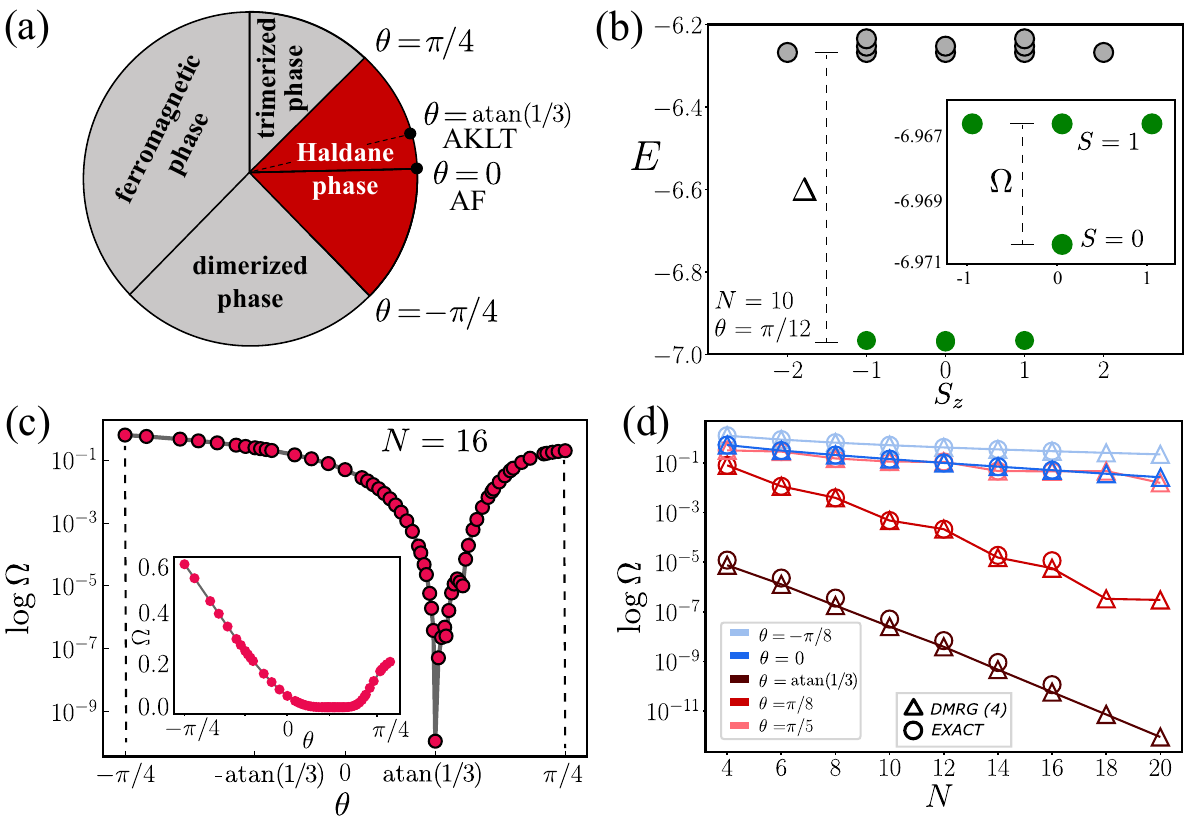}
	\caption{Energy properties of the spin $S=1$ bilinear-biquadratic Hamiltonian. (a) The phase diagram. (b) Energy spectrum vs. the total $S_z$ quantum number, for the system with $\theta=\pi/12$ and $N=10$. The energy splitting between the spin singlet and the spin triplet (green circles) is $\Omega=0.004$ while the energy separation between the four lowest energy states and excited ones is $\Delta=0.7$. (c) Dependence of the splitting $\Omega$ on the model, for a fixed chain length. (d) Exponential decay of $\Omega$ with chain length, for various models (defined by $\theta$) in the Haldane phase. The circle vs. triangle symbols indicate the numerical method used (see text).}
	\label{Figure1}
\end{figure*}

\paragraph*{Phase diagram and the gap of spin-1 chain models.---} We consider the family of spin-$1$ bilinear-biquadratic Hamiltonians for a chain of $N$ spins, which is conveniently parametrized by means of the angle $\theta$ as follows \cite{RenardRev}:
\begin{eqnarray}
H=J \sum_{j=1}^{N-1}\biggl[ \cos \theta \boldsymbol{S}_j \cdot \boldsymbol{S}_{j+1} +\sin \theta (\boldsymbol{S}_j \cdot \boldsymbol{S}_{j+1})^2\biggr],
\label{Hamiltonian}
\end{eqnarray}
where the terms proportional to $\cos \theta$ and $\sin \theta$ are, respectively, the nearest neighbour Heisenberg exchange interaction and the biquadratic exchange term. From now on we fix the energy scale as $J=1$.
Note that the Hamiltonian for any $\theta$ is invariant under global spin rotations, and hence eigenstates are in principle labeled by the total spin $S$ and total $S_z$ quantum numbers. The phase diagram of this Hamiltonian is well-known \cite{Nomura,RenardRev} and is conveniently captured by the angular diagram shown in Fig.~\ref{Figure1}(a). The Haldane phase, which occurs for $-\pi/4< \theta < \pi/4$, shows fractionalized $S=1/2$ states nucleating at the edges of the chain, arising from the topology of the Hamiltonian\cite{RenardRev}. Indeed, the low-lying energy levels are described by a spin singlet ($S=0$) and a spin triplet ($S=1$), separated by the energy $\Omega$, reaching a four-fold degeneracy in the thermodynamic limit \cite{PhysRevLett.111.167201,PhysRevB.51.16115,PhysRevB.53.40,PhysRevB.53.R492}. The higher energy states are separated from the low-energy ones by an energy gap $\Delta$. To illustrate that, we plot the energy spectrum of $H$  in Fig.~\ref{Figure1}(b) as a function of the eigenvalue $m$ of the total $S^z$ operator for $\theta=\pi/12$.\\ 
An exact solution for the ground state has been found for $\theta=\arctan(1/3)$ by Affleck, Kennedy, Lieb and Tasaki (AKLT chain) \cite{AKLT}. The AKLT ground state can be expressed as a symmetrical linear superposition of spin $1/2$ states where one of the two $S = 1/2$ variables at each site is paired in a (valence bond) singlet state with one of the two $S = 1/2$ variables at the neighbouring site. Due to this special configuration, the edge states of an open chain of any length are localized exactly on the end-sites, hence they are decoupled and exactly degenerate, so that the splitting $\Omega=0$. A similar picture emerges in the $\theta=0$ case as first pioneered  by Haldane \cite{PhysRevLett.50.1153}, but the localization length of the edge states, and hence their overlap and splitting energy $\Omega$, is finite away from the AKLT point. Fig.~\ref{Figure1}(c) shows that indeed there is a small range around the AKLT point where $\Omega(\theta)$ is strongly suppressed. We also confirm that the splitting $\Omega$ decays exponentially with the chain length $N$ for a generic $\theta$ in the Haldane phase, as expected\cite{PhysRevB.58.9248} from the exponential decay of the edge states, see Fig.~\ref{Figure1}(d). It is also clear that for a given chain the value of splitting energy $\Omega$ is suppressed by orders of magnitude as the model nears the AKLT point. Hence, beside the Haldane gap $\Delta$, the splitting energy $\Omega(\theta,N)$ is an informative observable in the Haldane phase.

Fig.~\ref{Figure1}(d) demonstrates the agreement between the lowest excitation energies obtained with two techniques (see \cite{SupMat}). First, numerical exact diagonalization, which we used for chains up to $N=16$. Second, having established the peak at $\Omega$, we used DMRG\cite{ITensor,ITensor-r0.3} for chains up to $N=30$. With DMRG we extract the four lowest energy states in the Haldane phase by finding the lowest state in the $S_z=\pm1$ sectors, and the two lowest states in the $S_z=0$ sector, hence spanning the low-energy subspace in the $S=0$ and $S=1$ sectors.

\paragraph*{Dynamical spin structure factor (DSF).---} When unpolarized electromagnetic waves are used for excitation, the standard ESR absorption spectrum $I(\omega) $ at frequency $\omega$ is given by the dynamical spin structure factor $\chi$, i.e., the imaginary part of the dynamical spin susceptibility\cite{jensen1991rare,RevModPhys.39.348,PhysRev.130.890}, $I(\omega) \sim -\text{Im} K^{spin}(\omega)\equiv\chi(\omega)$.\cite{PhysRevB.97.104411} Motivated by this, we consider a situation where locally at each chain site $j$ the ESR-STM signal\cite{DELGADO2021100625} is determined by the on-site DSF, namely $I_{j}(\omega) \sim-\text{Im} K^{spin}_j(\omega)\equiv\chi_j(\omega)$. In linear response theory, the latter quantity can be expressed at zero temperature as
\begin{eqnarray}
	\nonumber
	\chi_j(\omega)=&&-\text{Im} G^R_{S^+_{j} S^-_{j}}(\omega)=\\ \nonumber
	&&\frac{\pi}{p}\sum_{GS=1}^{p} \sum_n \biggl[|\bra{GS} S^+_{j}\ket{n}|^2 \delta(\omega+E_0-E_n)\\
	&&-|\bra{GS} S^-_{j}\ket{n}|^2 \delta(\omega+E_n-E_0)\biggr],
	\label{eq:Susceptibility}
\end{eqnarray} 
where $G^R$ is the retarded Green's function for the spin ladder operators $S^{\pm}=S^x\pm i S^y$ (see \cite{SupMat} for further details), while we used the Lehmann representation in the frequency domain $\omega$ \cite{bruus2004many}. 
In Eq. \eqref{eq:Susceptibility} the two sums extend, respectively, over all the $p$ states of the degenerate ground state manifold, and over all the $3^N$ many-body eigenstates of the open chain with $N$ sites. Considering both positive and negative values of $\omega$, by applying the time reversal operation we can deduce, in absence of external constant magnetic field, the oddness $\chi_j(\omega)=-\chi_j(-\omega)$.
Furthermore, all our chain models are invariant under spatial inversion with respect to the middle point, so that $\chi_j(\omega)=\chi_{N-j+1}(\omega)$.
\begin{figure*}
	\includegraphics[width=\textwidth]{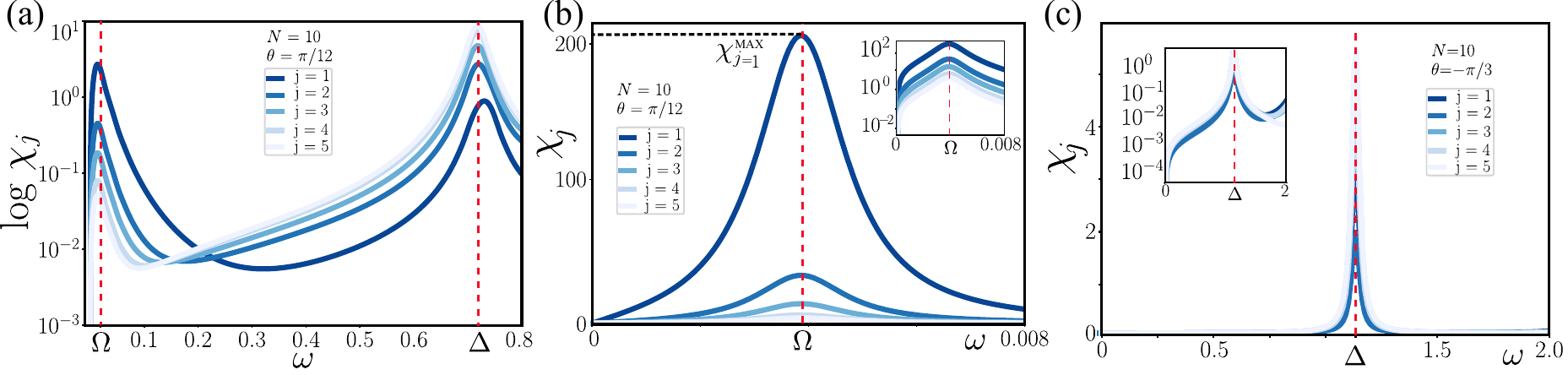}
	\caption{Frequency features of the atomically-resolved DSF $\chi_j(\omega)$. (a) Dependence on $\omega$ for a chain of $N=10$ sites in the Haldane phase, with $\Delta$ the Haldane bulkgap (units of $J$). The low-energy peak at the splitting energy $\Omega$ is characteristic of the topological edge states. We apply the usual Lorentzian broadening of the Dirac delta function (Eq.~\ref{eq:Susceptibility}) with width $\eta=0.02$. (b) A zoom in on the peak at $\Omega$, as panel (a) but with $\eta=0.001$. The inset shows the $\chi_j(\omega)$ on a log scale. (c) As panel (a) but for a chain in the trivial dimerized phase. There are no additional peaks below the one at the bulkgap.}
	\label{Figure2}
\end{figure*}

Figure~\ref{Figure2} illustrates our main observations about the DSF, contrasting the examples of an $N=10$ chain in the Haldane phase (Fig.~\ref{Figure2}(a),(b)), and in the topologically trivial dimerized phase (Fig.~\ref{Figure2}(c)). In both cases there is the dominant peak at the energy $\Delta$, which corresponds to the bulkgap, and is of the order of $J$. However, only in the Haldane phase there is an additional isolated peak at the low energy $\Omega$ caused by transitions between the $S=0$ and $S=1$ states spanned by the two spin-$1/2$ edgestates (compare Fig.~\ref{Figure2}(a) and (c)). Depending on the evenness or oddness of $N$, either the $S=0$ or the $S=1$ will be the lowest in energy, but the DSF $\Omega$ either way comes from the fact that there is an allowed transition. We note that for chains up to $N=10$ we obtain the entire spectrum by exact diagonalization, and perform the full sum over excited states $n$ in Eq.~\ref{eq:Susceptibility}. For longer chains we limit ourselves to states up to a cutoff energy $w\le 5\Delta$, and check that the DSF up to $1.5\Delta$ does not depend on the cutoff. The number of excited states increases smoothly as $\omega$ increases above the bulkgap, hence there are not too many states to include and we can use exact diagonalization for $N<18$ (see \cite{SupMat}). In the Haldane phase we now focus on the $\Omega$-peak. Fig.~\ref{Figure2} illustrates how the $\Omega$-peak decays monotonically with each site away from the edge. The $\Delta$-peak grows as we move away from the edge, and in the deep interior of the chain develops a flat profile characteristic of the bulk (not plotted). Fig.~\ref{Figure2}(b) focuses on the $\Omega$-peak. Note that the width of the peak is entirely due to the Lorentzian broadening we added.
\begin{figure}
	\includegraphics[width=0.47\textwidth]{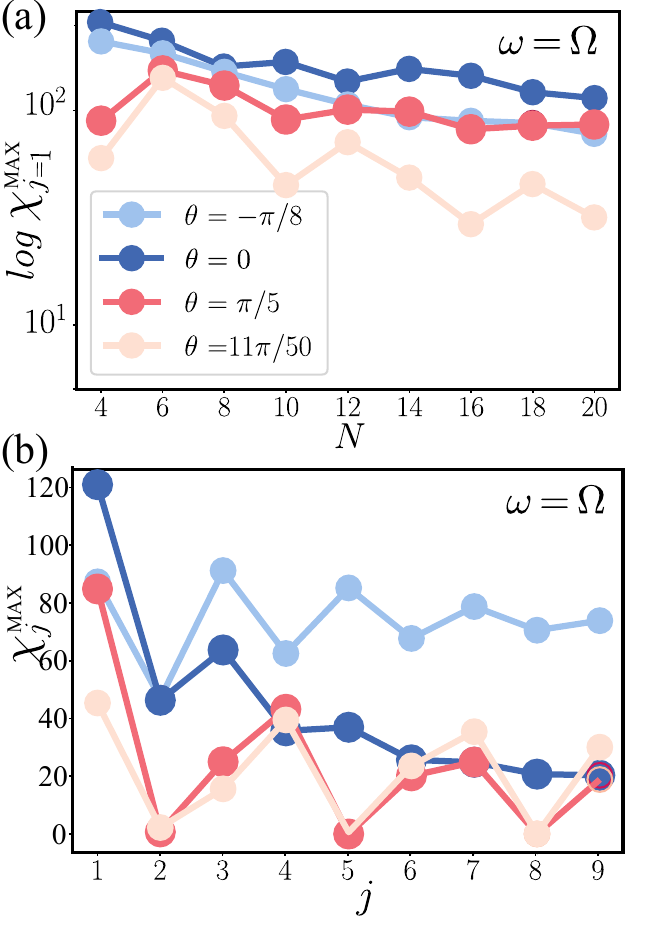}
	\caption{(a) The signal $\chi_{j=1}^{MAX}$ vs $N$ for four selected models, obtained by DMRG simulations. In DMRG, for the two states with total $S_z=\pm1$ we fix a bond dimension of $50$ and we perform $300$ sweeps, while for finding the two lowest statez with $S_z=0$ we use a bond dimension of $1200$ and we perform $600$ sweeps. (b) Scaling of $\chi_j^{MAX}$ vs $j$ for the same values of $\theta$ as in panel (a), for a chain of $N=18$ sites.}
 	\label{Figure3}
 \end{figure}

We further perform a systematic study of the DSF peak at energy $\Omega$, which characterizes the Haldane phase, by varying the model, chain length, and by observing site-by-site along the chain. We considered chains up to $N=30$ by DMRG, since only the 4 lowest states are necessary. We confirmed by exact diagonalization for shorter chains that adding the contribution of further excited states does not change the results\cite{SupMat}. In Fig.~\ref{Figure3}(a) we present the evolution with chain length of the strongest signal in the $\Omega$-peak of the Haldane phase, i.e., the quantity $\chi_{j=1}^{MAX}$, which is the peak value of the DSF achieved at $\omega=\Omega$, observed on the edge site $j=1$. The signal decays quite slowly, by less than a decade as the chain is quadrupled in length. This holds for models which are away from the AKLT point. At the AKLT point the value of $\Omega$ goes to zero, while the DSF is antisymmetric in $\omega$ and hence vanishes at $\omega=0$ (see after Eq.~\ref{eq:Susceptibility}), so the height of the peak $\chi_{j=1}^{MAX}$ has to vanish at the AKLT point and is suppressed for models near it (see \cite{SupMat} for details). Quantitatively, as long as the chain is not too long, and the model is not too close to AKLT, so that the peak position $\Omega/J > 5*10^{-3}$, the peak height $\chi_{j=1}^{MAX}$ remains numerically non-zero. These properties suggest that antiferromagnetic Heisenberg spin chains of moderate lengths below $N=30$ unambiguously distinguish the Haldane phase by means of the DSF. (For longer chains, an external magnetic field can be used to circumvent the issue of vanishing $\Omega$ (see below)).

In Fig. \ref{Figure3}(b) we show the decrease of the signal $\chi_{j=1}^{MAX}$ when $j$ moves from an edge of the chain towards the middle point, for several values of $\theta$ inside the Haldane phase. Note, the spin-spin correlation function also decays roughly exponentially (deviations are expected before the asymptotic regime is reached\cite{Polizzi}) but peaks again at the opposite end of chain due to the overlap of edge states (see \cite{SupMat}). Apart from the expected decay from the edge towards the middle of chain, we observe oscillations with period two for $\theta<0$, i.e., for the part of Haldane phase nearer to the dimerized phase, and with period three for $\theta>0$, i.e., the part of Haldane phase nearer to the trimerized phase. These oscillations have not been analyzed in detail\cite{Polizzi}. In longer chains, the decay is more pronounced, but the oscillations become more regular and period two dominates, so that by length $N=28$ we cannot identify the period three for the studied $\theta$ values. Observing the atomically resolved DSF at the $\Omega$ peak could therefore reveal the sign of $\theta$ within the Haldane phase, but only in moderately long chains.

\paragraph*{Tuning by external magnetic field.---} We consider a constant uniform external magnetic field acting as a Zeeman term added to the Hamiltonian in Eq.~\ref{Hamiltonian}:
\begin{equation}
H_B=B\sum_{j=1}^{N}S_j^z=B S^z,
\label{eq:Zeeman}
\end{equation}
where $S^z$ is the total spin-$z$ component, which is still conserved by $H+H_B$. Hence, the Zeeman field simply splits the $S=1$ triplet part of Haldane edge states to energies $0,\pm B$. The DSF (Eq.~\ref{eq:Susceptibility}) with $S=0$ the lowest-energy edge state component now contains one peak at $\omega_+=\Omega-B$ from the spin-lowering transition to $S=1,S^z=-1$, and another peak at $\omega_-=-(\Omega+B)$ from the spin-increasing transition to $S=1,S^z=1$. The breaking of time-reversal symmetry now obviously breaks the antisymmetry of DSF as peaks are not even at opposite energies, $\omega_-\neq-\omega_+$. Hence, the signature peak at $\Omega$ and its equivalent at $-\Omega$ are not split, but shifted uniformly by $B$, at least while $|B|<\Omega$ and no reordering of states occurs (see \cite{SupMat} for details). The magnetic field hence allows a simple tuning of a too-small $\Omega$ in long chains so as to make the peak observable. We also confirmed that the spatial profile of the DSF signal is not affected by the magnetic field (see \cite{SupMat}).

\paragraph*{Robustness to anisotropy.---} For chains grown on surfaces for access by ESR-STM, in principle the substrate induces spin anisotropy. We hence consider the effect of adding a uniaxial single-ion anisotropy
\begin{equation}
H_D=D\sum_{j=1}^{N} (S_j^z)^2,
\end{equation}
to the chain Hamiltonian, Eq.~\ref{Hamiltonian}. This anisotropy is among the most important perturbations in real chains, but it is known that the Haldane phase survives with a reduced gap $\Delta$ even for quite large values $-0.3<D/J<1$. Here we consider small but non-negligible perturbations $D<0.1$ to stay within the Haldane phase. Regarding the four low-energy levels, the anisotropy splits the $S=1$ triplet into $S_z=0$ and $S_z=\pm1$. The anisotropy splitting $\delta_D$ is small, of order $\delta_D\sim D^2$ for small $D$ (see \cite{SupMat} for more details). In principle this may cause a new low-energy peak in DSF at $\omega=\delta_D$, but for our range of values of $D$ this is such a small scale that it is suppressed by the fact that DSF is zero at $\omega=0$ (see below Eq.~\ref{eq:Susceptibility}. On the other hand, the height of the DSF peak at $\omega=\Omega$ is more strongly influenced by the anisotropy, surprisingly it increases by about $10\%$ for $D/J=0.1$, although it does not seem to affect the spatial behavior of the DSF peak (see \cite{SupMat}). We conclude that a weak anisotropy does not affect our qualitative conclusions about the DSF in the Haldane phase, and it may even be beneficial for detection of the $\Omega$ peak.

\paragraph*{Discussion and conclusions.---} In conclusion, we have established the local dynamical spin structure factor as a useful probe of the Haldane phase of spin-$1$ antiferromagnetic chains. The probe utilizes both the universal topologically protected ground state manifold, and its non-universal splitting in energy by the finite chain length. We find that for a large range of chain lengths and a wide range of Hamiltonian parameters in the Haldane phase of the bilinear biquadratic Hamiltonian, the peak in DSF at the energy of the splitting appears as an observable. We don't expect these conclusions to be changed by considering other spin models exhibiting the Haldane phase. Experimentally, considering $S=1$ chains\cite{Mishra2021} based on open-shell nanographene molecules, the typical exchange scale is $J\sim1-10meV$, while the energy scales best probed by ESR-STM are of order $\omega>40\mu eV$, so one would prefer a system where the signature of the Haldane phase is at $|\Omega\pm B|/J\sim1/20$. Further, for carbon- and molecule-based spin systems, one expects a dominant standard antiferromagnetic coupling, $\theta=0$. Finally, in contrast to magnetic adatoms, the single-ion anisotropy is very small, $D/J<10^{-3}$. Hence we predict that without magnetic field optimal chain lengths are $N=10\div30$, while for longer chains a magnetic field of order Tesla is enough to make the peak observable. Our calculations are at zero temperature. A finite temperature that is less than the Haldane gap, $k_B T\ll\Delta$ would start populating both $S=0$ and $S=1$ low-lying states, and hence would diminish the intensity of the signature peak. In presence of magnetic field, a temperature of order $min(\Omega,|\Omega-B|)$ would introduce new peaks at exact energies $\pm B$ from transitions within the $S=1$ sector, but it wouldn't introduce any other features from states beyond the bulkgap. Given that in Haldane chains of open-shell nanographene and porphyrin molecules the typical $J=1-10meV$, the thermal broadening hence should stay well below the scale of $50K$ to prevent non-topological peaks from beyond the bulkgap. As a further example, for chains of order $N=10$ the $\Omega\sim0.1\Delta$, so the temperature should be further reduced below $\sim 5K$ to avoid diminishing the peak, while for chains above $N=30$ where $\Omega$ becomes too small, the field can be tuned to about a Tesla and the temperature kept at about a Kelvin. Experimentally, the main perturbation is expected to be an anisotropy, but for adsorbed molecules it is quite small, and we show that hence it doesn't affect our findings. On the other hand, an open challenge remains to consider other dynamical observables which may be more accurate in describing the experimental ESR-STM signal.
\begin{acknowledgments}
\paragraph*{Acknowledgments.---}A.M., P.S. and H.A. acknowledge the support of the French Agence Nationale de la Recherche (ANR), under grant number ANR-22-CE30-0037.
\end{acknowledgments}

\bibliography{BibHaldane}

\end{document}


\title{Supplementary Material for "Local dynamics and detection of topology in spin-1 chains"}

\author{Alfonso Maiellaro}
\affiliation {Université Paris-Saclay, CNRS, Laboratoire de Physique des Solides, 91405 Orsay, France}
\affiliation {INFN, Sezione di Napoli, Gruppo collegato di Salerno, Italy}
\author{Herv\'e Aubin}
\affiliation {Universit\'e Paris-Saclay, CNRS, Centre de Nanosciences et de Nanotechnologies, 91120, Palaiseau, France}
\author{Andrej Mesaros}
\affiliation{Université Paris-Saclay, CNRS, Laboratoire de Physique des Solides, 91405 Orsay, France}
\author{Pascal Simon}
\affiliation{Université Paris-Saclay, CNRS, Laboratoire de Physique des Solides, 91405 Orsay, France}

\date{\today}

\maketitle


\section{Density of states and summing over excited states}
\begin{figure}
 		\includegraphics[scale=0.55]{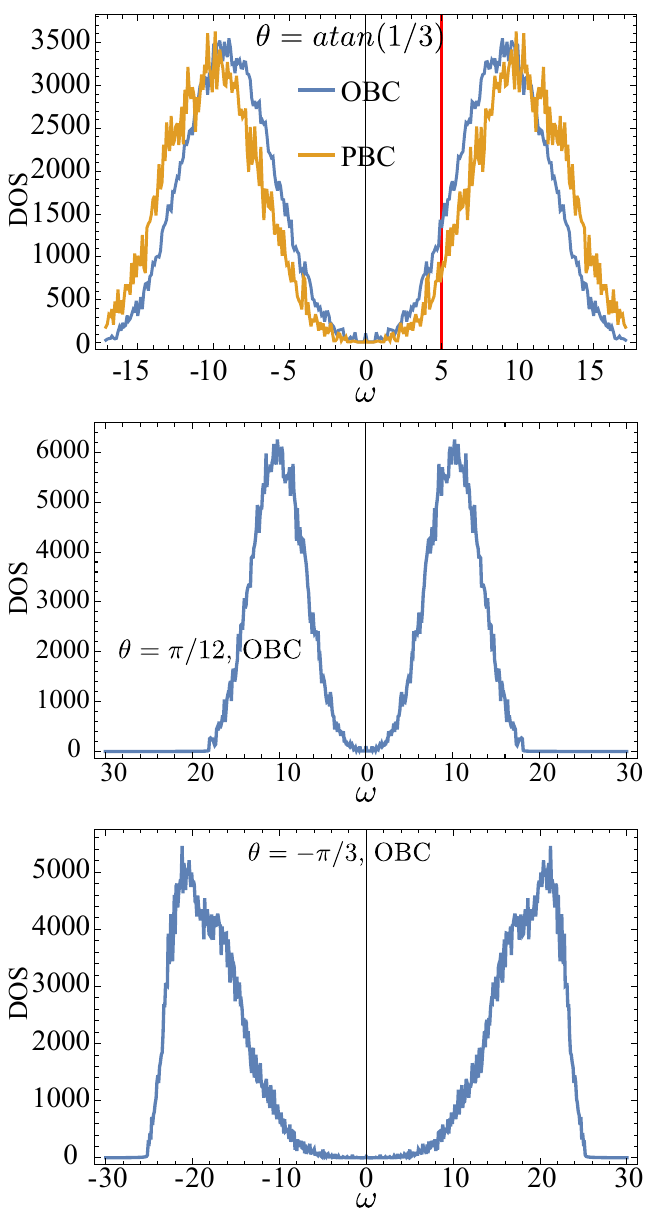}
	\caption{Density of states (symmetrized in $\omega$) for a $N=10$ chain with broadening $\eta=0.02$, representing the entire spectrum obtained by exact diagonalization. (a) AKLT point, periodic (PBC) and open (OBC) boundary conditions for the chain. The red vertical line marks our cutoff $\Lambda=5J$ for calculating the $\chi_j(\omega)$. The bulkgap is $\Delta\approx0.7J$. (b) Haldane phase, $\theta=\pi/12$, open chain. (c) Dimerized phase, $\theta=-\pi/3$, open chain.}
	\label{SuppDOS}
\end{figure}
The density of states for gapped phases we consider has the same feature: it starts raising slowly from zero as $\omega$ passes the bulkgap $\Delta\sim J$, see Fig.~\ref{SuppDOS}. Hence there are not too many states at energies of order a few $\Delta$, which helps the calculation of the DSF. Namely, as we calculate at zero temperature and are interested in DSF feature up to $\omega\sim\Delta$, we only have to sum excited states up to that energy. In practice, we use the high-energy cutoff $\Lambda=5J$ for the sum (see Fig.~\ref{SuppDOS}(a)), and check that the results do not change when varying $\Lambda$. Finally we note that the total range of energies in the spectrum increases with $N$.

\section{Spin-spin correlation function}
\begin{figure}
	\includegraphics[width=0.5\textwidth]{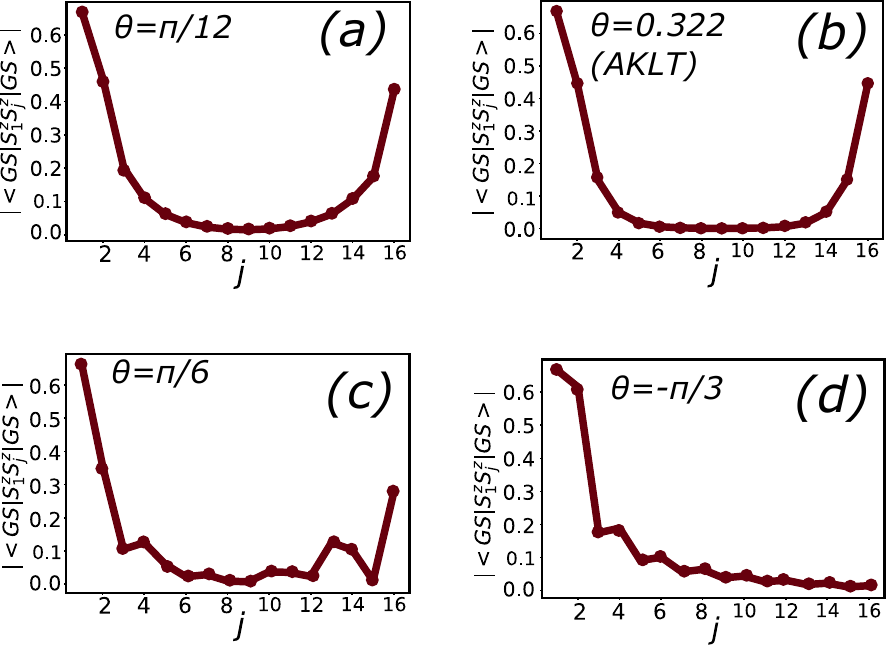}
	\caption{Absolute value of the correlation function from the left edge, $|\bra{GS}S^z_1 S^z_j\ket{GS}|$, for a chain of $N=16$ sites. (a)-(c) Inside the Haldane phase, there are non-local correlations between the edges. (d) Outside the Haldane phase, the correlation just decays exponentially.}
	\label{Supp1}
\end{figure}

Inside the Haldane phase (Fig.~\ref{Supp1}(a)-(c)), the site dependence of the correlation function from the first site $|\bra{GS} S^z_1 S^z_j \ket{GS}|$ shows an exponential decay towards the middle of the chain, and then an increase so that there is a second peak at the opposite end of the chain. This reveals the correlations due to the overlapping edge states. In accord, the peak at the opposite end of the chain ($j=N$) diminishes as we approach the phase transition out of the Haldane phase. Outside the Haldane phase (Fig.~\ref{Supp1}(d)), the $|\bra{GS} S^z_1 S^z_j \ket{GS}|$ is peaked at one edge ($j=1$) and vanishes exponentially all the way towards the other edge. Note, due to spin-rotation invariance of the Hamiltonian in Eq.~1, the correlation function $\bra{GS} \vec{S}_i\cdot \vec{S}_j \ket{GS}=3\bra{GS} S^z_i S^z_j \ket{GS}$.

\section{Additional results on the dynamic spin structure factor}
\begin{figure}
	\includegraphics[width=0.45\textwidth]{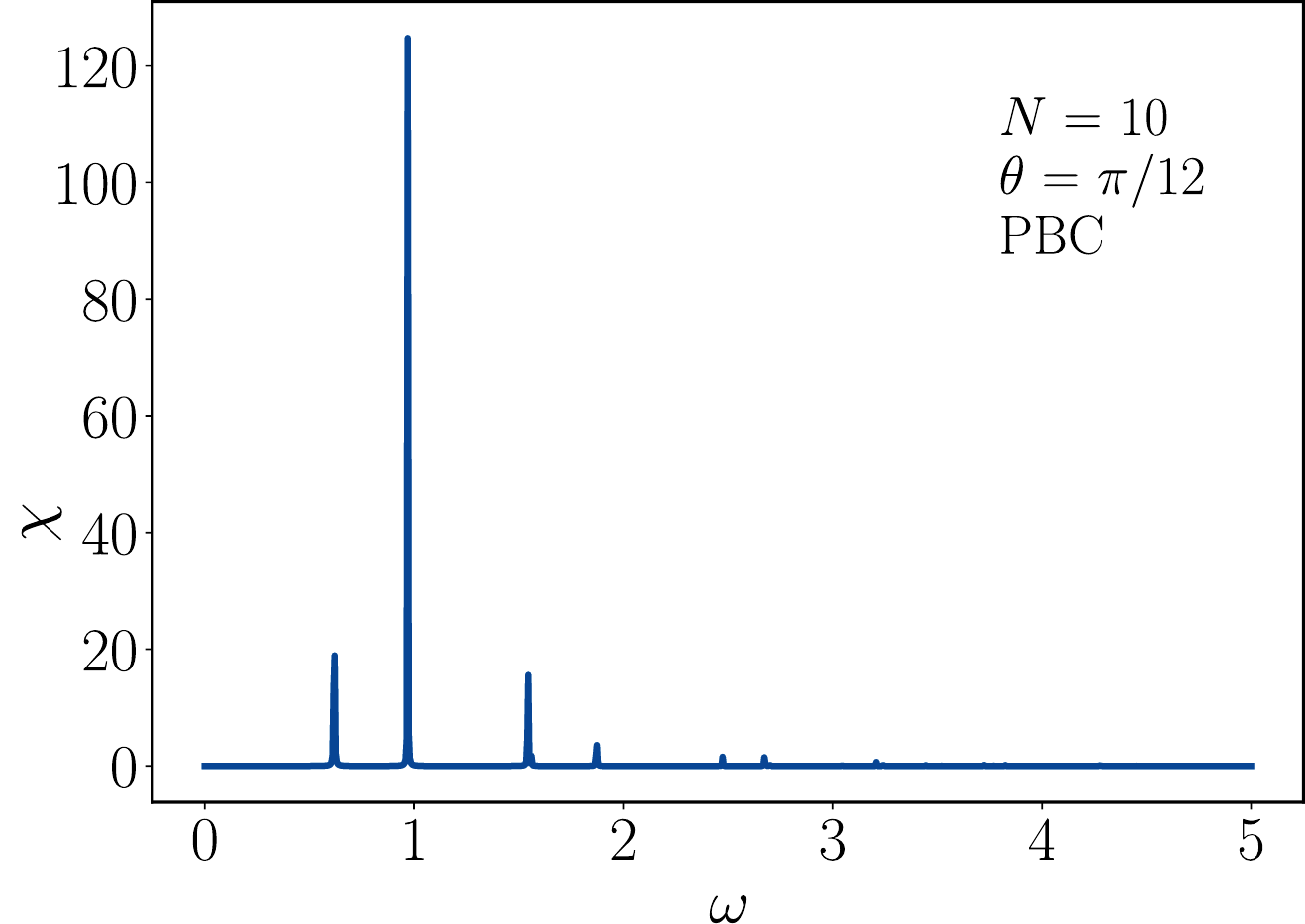}
 \caption{The DSF on a site of a periodic chain in the Haldane phase, on a wider scale of energies up to $\omega=5J$. Maximal energy for this chain is $20J$ (see Fig.~\ref{SuppDOS}(b)).}
	\label{FigSupPBC}
\end{figure}
In Fig.~\ref{FigSupPBC} we show the DSF for a periodic chain in the Haldane phase, demonstrating the absence of low-energy peaks due to the absence of (quasi)degenerate edge states. We show a wider range up to $\omega=5J$ to demonstrate the sparse peak structure beyond the bulkgap $\omega=\Delta\sim0.7J$. Note that the majority of the $3^10$ states in the spectrum lay at the even higher energies up to the maximal $\omega=20J$ (see Fig.~\ref{SuppDOS}(b)), but the peak structure remains sparse due to the strong selection rule $\delta S_z=\pm1$ in the DSF.

\begin{figure}
 		\includegraphics[width=0.45\textwidth]{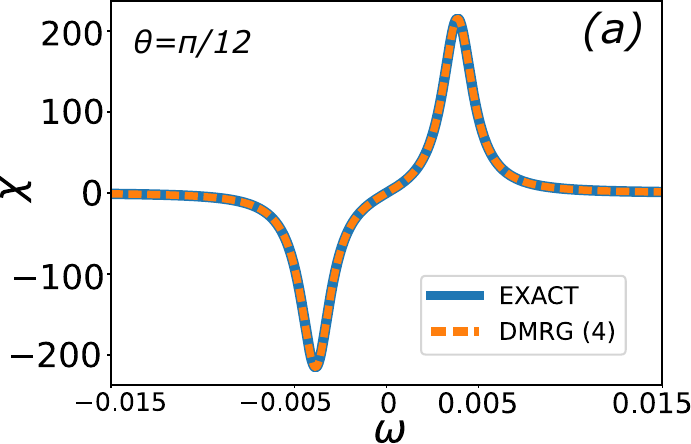}\\
 		\vspace{0.5cm}
	\includegraphics[width=0.45\textwidth]{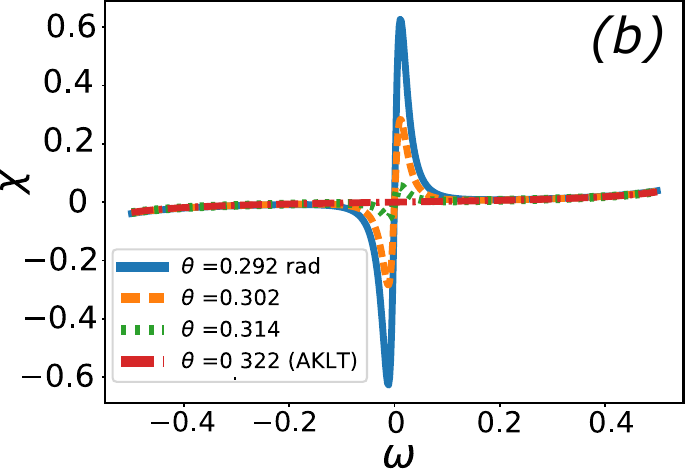}
	\caption{(a) Comparison between the DSF $\chi_{j=1}(\omega)$ obtained by exact diagonalization and by DMRG, for an edge site at low energies $\omega\lesssim\Omega$. The DMRG DSF has been obtained by only including the four lowest energy states.  In the DMRG simulations, in the sector $S_z=0$ we fix for the states $\ket{0\ 0}$ and $\ket{1\ 0}$ a bond dimension of $1200$ and we perform $600$ sweeps, while for states $\ket{1\ 1}$ and $\ket{1\ -1}$ we fix a bond dimension of $50$ and we perform $300$ sweeps. (b) $\chi$ near the AKLT limit. For both the panels we are considering a chain of $N=10$ sites.}
	\label{Supp2}
\end{figure}

In Fig. \ref{Supp2} (a) we compare the DSF $\chi$ obtained by exact diagonalization with $\chi$ obtained by means of the DMRG simulations. The exact $\chi$ has been obtained by summing over all the $3^N$ states while the DMRG $\chi$ has been obtained only including the lowest four energy states $\ket{S\ S_z}=$ $\ket{0\ 0}$, $\ket{1\ 0}$, $\ket{1\ 1}$, $\ket{1\ -1}$. We have numerically verified that near the $\Omega$ frequency, $\chi$ is entirely due to these four states. The quantitative matching of the two plots also shows the convergence of the DMRG simulations.

In Fig. \ref{Supp2} (b) we investigate $\chi$ near the AKLT point. Exactly at the AKLT point,  $\theta=\arctan 1/3$, the $\Omega$ peak is absent (red curve) because of the exact degeneracy of the four lowest states, already present for short chains. Nevertheless a small shift of model, $\delta\theta=0.012$, is already sufficient to restore the $\Omega$ peak in $\chi$ (green and orange curves).

\begin{figure}
 		\includegraphics[width=0.45\textwidth]{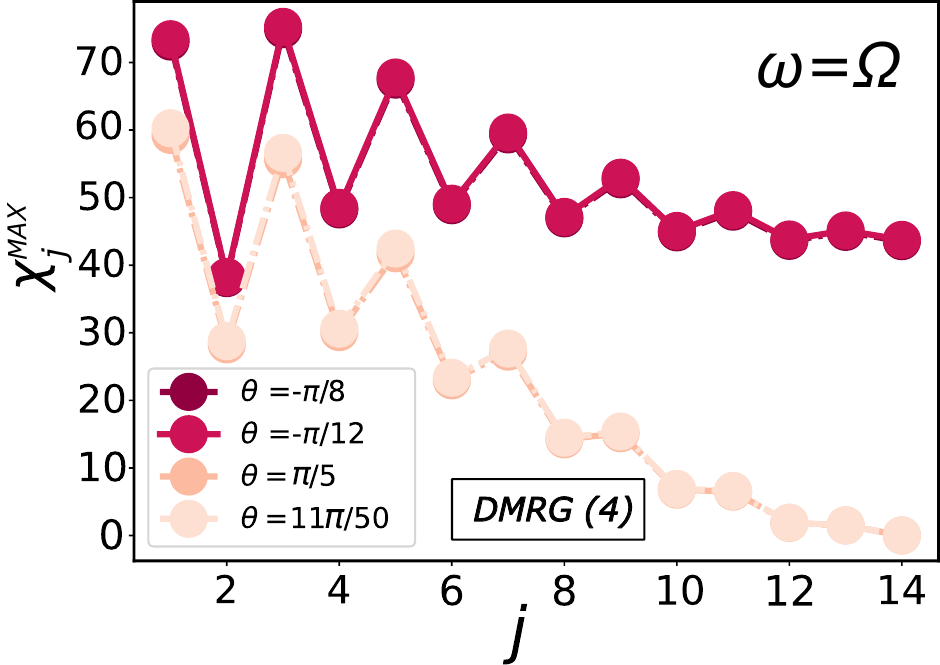}
	\caption{Scaling of $\chi_j^{MAX}$ vs $j$ for the same values of $\theta$ as in panel Fig.3(b) of main text, but for a longer chain of $N=28$ sites.}
	\label{FigSupN28}
\end{figure}

Finally, in Fig.~\ref{FigSupN28} we present the DSF signal site-by-site on longer chains $N=28$ throughout the Haldane phase, analogous to the Fig.3(b) of main text which is te same plot for $N=18$ chains. One can see that for long chains the oscillations develop the same periodicity of 2 sites, even when the Hamiltonian is close to the transition towards the trimerized phase.

\section{Magnetic field dependence}
\begin{figure}
 		\includegraphics[width=0.45\textwidth]{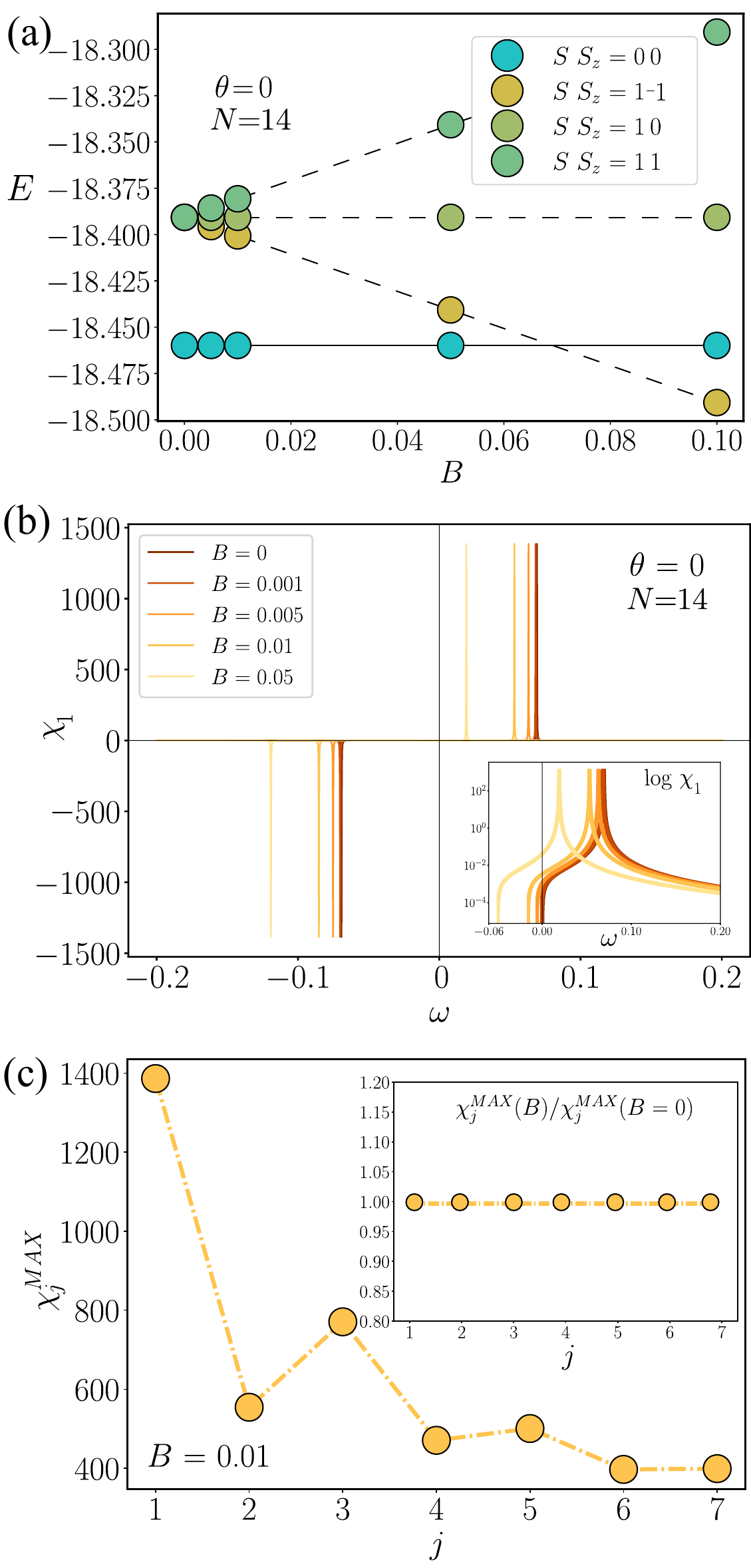}
	\caption{\label{FigSupB}Example of effect of uniform Zeeman field on a Heisenberg chain in Haldane phase. (a) The linear splitting by $B$ of edge states due to Zeeman field. (b) The DSF as function of both positive and negative $\omega$, showing the linear shift of peaks by $B$. The peak at $\omega_+=\Omega-B$ (in this example $\omega_+>0$, and we chose $B>0$) is caused by a spin-lowering transition from the $S=0$ ground state, while the peak at $\omega_-=-\Omega-B$ (here $\omega_-<0$) is due to the spin-raising transition. (c) The DSF as function of site $j$ on chain, at the maximum signal $\omega=\omega_+$. The inset confirms that the spatial profile of DSF didn't change compared to the signal at $\omega=\Omega$ for $B=0$.}
\end{figure}
Fig.~\ref{FigSupB} shows an example of how a uniform magnetic field shifts the DSF peak in energy, without affecting other properties, at least while $|B|<\Omega$. We note that both here and in the main text we presented examples of positive Zeeman field $0<B<\Omega$, however the general peak structure and dependence on $\omega_\pm=\pm\Omega-B$ remains for any sign of $B$, as long as $|B|<\Omega$. We also only presented even-length chains for which the lowest-energy component of edge states is the $S=0$. In odd-$N$ chains the order of states at $B=0$ is reversed~\cite{PhysRevLett.111.167201}. For odd-$N$ the spin-lowering or raising transition would still be at $\omega_+=\Omega-B>0$, and $\omega_-=-\Omega-B$, respectively, where $\Omega>0$ is now the $B=0$ splitting between the $S=1$ ground state and the excitation with $S=0$. In odd chains, even when $|B|>\Omega$, the ground state would remain the same ($S_z=\mp1$ for $B>0(B<0)$) and nothing changes as the peaks are rigidly shifted by $B$. For even-$N$ chains the evolution is more complicated. Once $|B|>\Omega$, the ground state changes from $S=0$ to $S_z=\mp1$ in case of $B>0(B<0)$, see Fig.~\ref{FigSupB}(a). Focusing on $B>0$, as $B$ becomes greater than $\Omega$, and the $\omega_+=\Omega-B$ changes sign, the spin-lowering transition ($S=0$ to $S_z=-1$ at $\omega_+>0$) changes to a spin-raising one ($S_z=-1$ to $S=0$ at $\omega_+<0$), while the spin-raising one ($S=0$ to $S_z=+1$ at $\omega_-=-\Omega-B<0$) is discontinuously replaced by another spin-raising transition $S_z=-1$ to $S=1,S_z=0$ at $-B<0$. Hence in this example of even-$N$ with $B>0$ at zero temperature, a positive DSF peak at $\omega=\omega_+=\Omega-B>0$ passes through $\omega=0$ (at the tuning $|B|=\Omega$) and emerges as a negative DSF peak at $\omega=\omega_+=\Omega-B<0$, while the negative DSF peak at $\omega=\omega_-=-\Omega-B<0$ is discontinuously replaced by also a negative DSF peak but at $\omega=-B<0$, hence at the tuning $|B|=\Omega$ this peak's position jumps from $\omega=-2B=-2\Omega$ to $\omega=-B=-\Omega$.

\section{Single-ion anisotropy}
\begin{figure*}[h!]
	\includegraphics[width=\textwidth]{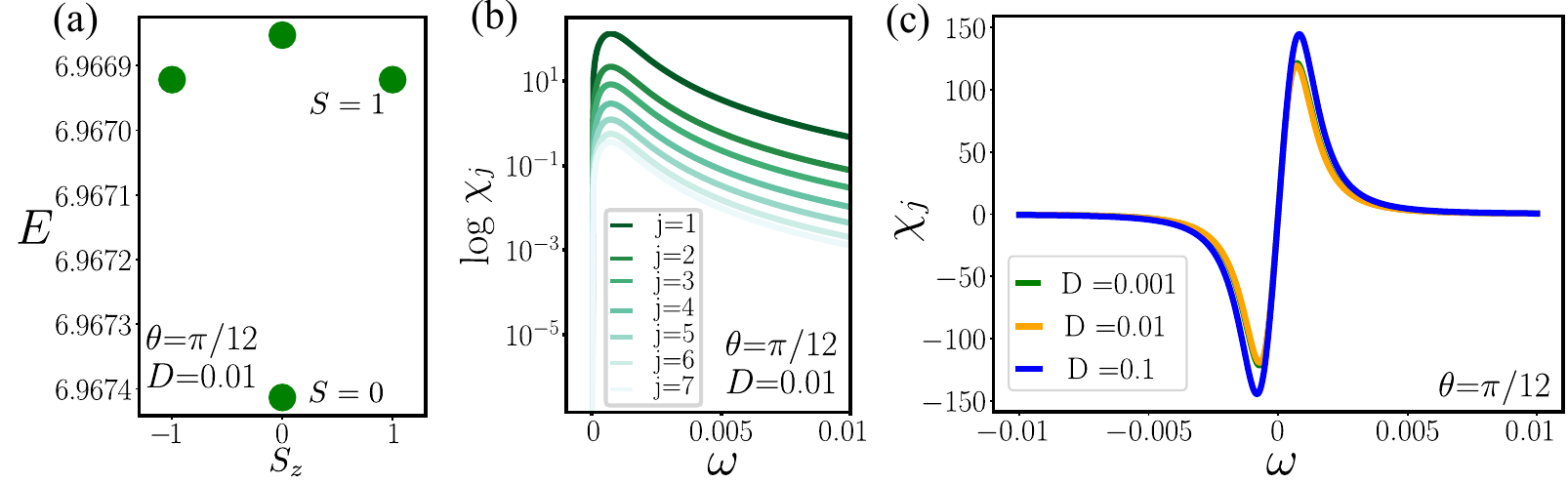}
	\caption{\label{FigSupAnis}Effect of single-ion spin anisotropy of strength $D$ on a $N=14$ chain in the Haldane phase. (a) The splitting of the $S=1$ sector of the low-energy states. (b) The dependence of the tomographic DSF on $\omega$. (c) Dependence on D of the DSF at the edge site.}
\end{figure*}

Fig.~\ref{FigSupAnis}(a) illustrates how the anisotropy splitting $\delta_D$ is of order $\delta_D\sim D^2$ for small $D$. The slight increase of the height of the DSF peak at $\omega=\Omega$ due to the anisotropy is presented in Fig.~\ref{FigSupAnis}(c)), while the spatial behavior of the DSF peak semmes less affected (Fig.~\ref{FigSupAnis}(b)).

\bibliography{BibHaldane}